\documentclass[journal]{rmaa}

\usepackage{paralist}

\usepackage{psfrag,color}

\usepackage[latin1]{inputenc}

\title{Chemical abundances for A-and F-type supergiant stars} 

\author{
  R. E. Molina,\altaffilmark{1} 
  and H. Rivera,\altaffilmark{1}}

\altaffiltext{1}{Laboratorio de Investigaci\'on en F\'{i}sica Aplicada y Computacional,
  Universidad Nacional Experimental del T\'achira, CP 5001, Venezuela.}

\shortauthor{Molina \& Rivera}
\shorttitle{Elemental abundances in supergiants}

\fulladdresses{
\item R. E. Molina: Laboratorio de Investigaci\'on en F\'{i}sica Aplicada y
Computacional, Universidad Nacional Experimental del T\'achira,  Venezuela,
(rmolina@unet.edu.ve). 
\item H. Rivera: Laboratorio de Investigaci\'on en F\'{i}sica Aplicada y
Computacional, Universidad Nacional Experimental del T\'achira,  Venezuela,
(hrivera@unet.edu.ve).}
\listofauthors{R. E. Molina \& H. Rivera}
\indexauthor{Molina, R. E.}
\indexauthor{Rivera, H.}
\abstract{We present the stellar parameters and elemental abundances of 
a set of A--F-type supergiant stars HD\,45674, HD\,180028, HD\,194951 and 
HD\,224893 using high resolution ($R$\,$\sim$\,42,000) spectra taken from 
ELODIE library. We present the first results of the abundance analysis 
for HD\,45674 and HD\,224893. We reaffirm the abundances for HD\,180028 
and HD\,194951 studied previously by Luck (2014) respectively. 
Alpha-elements indicates that objects belong to the thin disc population. 
From their abundances and its location on the Hertzsprung-Russell diagram 
seems point out that HD\,45675, HD\,194951 and HD\,224893 are
in the post-first dredge-up (post-1DUP) phase and they are moving
in the red-blue loop region. HD~180028, on the contary, shows typical 
abundances of the population I but its evolutionary status could not 
be satisfactorily defined.}

\resumen{Hemos efectuado un an\'alisis detallado de las abundancias 
qu\'{i}micas de cuatro objetos supergigantes HD\,45674, HD\,180028, HD\,194951 
y HD\,224893, usando espectros de alta resoluci\'on ($R$\,$\sim$\,42,000) 
tomados de la libreria ELODIE. Se presentan los primeros resultados de las 
abundancias para HD\,45674 y HD\,224893, y se reafirman las abundancias para 
HD\,180028 y HD\,194951 calculadas por Luck (2014). Los elementos alfa nos 
indican que todos los objetos estudiados pertenecen a la poblaci\'on del disco 
gal\'actico. A partir de sus abundancias y de su localizaci\' on en el diagrama
Hertzsprung-Russell parece indicarnos que HD\,45675, HD\,194951 y HD\,224893 
evolutivamente se encuentran en la fase posterior al primer dragado (post-1DUP) 
y se mueven en la regi\'on del red-blue loop. HD~180028 muestra abundancias 
t\'{i}picas de la poblaci\'on I pero su estado evolutivo no puede ser definido 
satisfactoriamente.}

\addkeyword{Stars: atmospheric parameters}
\addkeyword{Stars: abundances}
\addkeyword{Stars: evolution}
\addkeyword{Stars: supergiant}

\begin{document}
% Typeset article header
\maketitle

\section{Introduction}
\label{sec:introd}

The process of chemical evolution in the Galaxy can be understood from its massive stars.
These objects in their rapid evolution undergo changes through the process of nucleosynthesis
over time and return their chemical elements into the interstellar medium from stellar winds
and supernova events. It is not surprise the existence of massive young objects in
the galactic plane since it is an area of star formation. These objects are visually
luminous in galaxies and, in general, suitable candidates for studies of stellar
and chemical evolution (Luck et al.\@ 1998; Smiljanic et al.\@ 2006; Venn et al.\@ 2000, 2001,
2003; Kaufer et al.\@ 2004).

In the galactic disc, some of these massive objects have been classified as supergiant stars
with masses between 5 to 20\,M$_{\odot}$, with A-and-F spectral type, which are moderately evolved
and where the chemical abundances of the light elements CNO have been key to discriminate their
evolutionary states (Lyubimkov et al.\@ 2011, Venn 1995a,1995b and internal references). For massive
stars when H is exhausted, the post-He core burning phase can be affected in several ways.

Stellar evolutionary models constructed at solar metallicities predict that massive supergiants
($M$\,$\geq$10\,M$_{\odot}$) are in the phase of helium core burning (Schaller et al.\@ 1992; Stothers \& Chin
1991). These objects have already left the main sequence on the Hertzsprung-Russell diagram
(HRD) and begins the
ignition of He in the blue supergiant region but thermal instabilities causes in the star a rapid
expansion towards the red supergiant one. In this last phase, the A--F supergiants are able to resume thermal
and radiative equilibrium through convection in the outer layers and shows an altered CNO
abundances due to the first dredge-up (1DUP) event.

On the other hand, A--F supergiants less massive
($M$\,$<$\,10\,M$_{\odot}$) have also initiated the He-core burning without visiting the red giant branch, a
fully convective intermediate zone is predicted, the envelope is able to establish the thermal
equilibrium and the stars are kept in the blue supergiant phase. Under these conditions CNO abundances
remain unchanged (Stothers \& Chin 1976; 1991).

However, another scenario is possible for objects
with intermediate masses (3\,$<$\,M\,$<$\,9\,M$_{\odot}$) and is called the blue loop.
In this point the star has eventually developed a convective envelope and is rising up
the Hayashi line in the HRD. These objects have already reached the red supergiant stage but
eventually evolved back into a blue supergiant phase (Walmswell et al.\@ 2015).
During the red supergiant phase, the
convective zone mixing materials of H-burning shell which are subsequently released to the surface
by the event of the 1DUP and it show changes in the observed CNO abundance patterns.
The amount CNO-processed material in such objects allows to discriminate different
types of supergiants.

The main goal of this work is based on a detailed study of chemical abundances for a set of
four low-latitude A--F supergiants HD\,45674, HD\,180028, HD\,194951 and HD\,224893
under LTE assumption
and the determination of their atmospheric parameters from excitation and ionization equilibrium.
For this purpose we employ high-resolution spectroscopy  and a set of
atmospheric models
constructed with plane-parallel geometry, hydrostatic equilibrium, local thermodynamic
equilibrium (LTE) and the \texttt{ODFNEW} opacity distribution (Castelli \& Kurucz 2004).
It is expected that the abundances derived correspond to the typical abundances observed for
supergiant stars in the galactic disc.

This paper are organized as follows: In \S~\ref{sec:observ} regards to the description of the sample
selection. In \S~\ref{sec:param} presents how the atmospheric parameters were estimated. We determine
the chemical abundances for the three stars in \S~\ref{sec:abund}. An individual analysis of
abundances for each stars is present in \S~\ref{sec:discuss}. In \S~\ref{sec:results} is dedicated to 
discuss our results, and finally, in \S~\ref{sec:concluss} gives the conclusions of the paper.

\begin{figure}
\centering
\includegraphics[width=7.5cm,height=7.5cm]{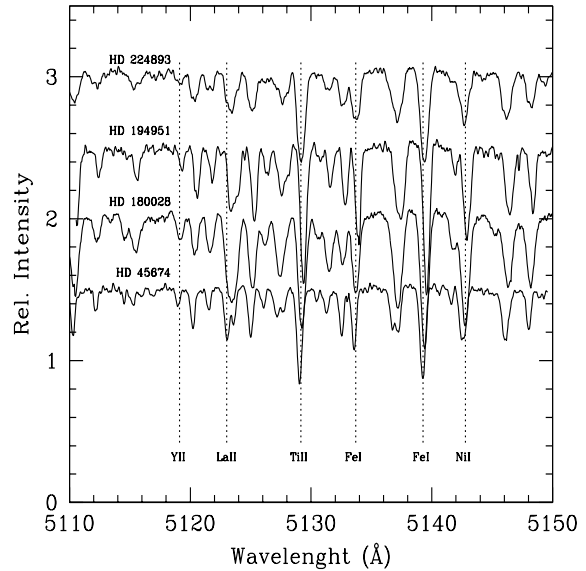}
\caption{Representative spectra of the sample stars HD\,45674, HD\,180028, HD\,194951
and HD\,224893. The
location of lines of certain important elements have been indicated by dashed lines.
This stars are arranged in decreasing order the HD number.}
\label{fig:figure1}
\end{figure}

\begin{table*}
\begin{center}
\begin{minipage}{170mm}
  \caption{Basic parameters for the sample.}
 \label{tab:table1}
\scriptsize{\begin{tabular}{rccccccccccc}
  \hline
  \hline
\multicolumn{1}{l}{No. HD}&
\multicolumn{1}{l}{SpT}&
\multicolumn{1}{c}{$\alpha$}&
\multicolumn{1}{c}{$\delta$}&
\multicolumn{1}{c}{$V$}&
\multicolumn{1}{c}{$B$}&
\multicolumn{1}{c}{$l$}&
\multicolumn{1}{c}{$b$}&
\multicolumn{1}{c}{$V_\mathrm{r}$}&
\multicolumn{1}{c}{$\pi$}&
\multicolumn{1}{c}{$E(B-V)$}&
\multicolumn{1}{c}{$v~sini$}\\
\multicolumn{1}{c}{}&
\multicolumn{1}{c}{}&
\multicolumn{1}{c}{(h~m~s)}&
\multicolumn{1}{c}{($^{o}$~$^{'}$~$^{"}$)}&
\multicolumn{1}{c}{(mag)}&
\multicolumn{1}{c}{(mag)}&
\multicolumn{1}{c}{($^{o}$)}&
\multicolumn{1}{c}{($^{o}$)}&
\multicolumn{1}{c}{(km\,s$^{-1}$)}&
\multicolumn{1}{c}{(mas)}&
\multicolumn{1}{c}{(mag)}&
\multicolumn{1}{c}{(km\,s$^{-1}$)}\\
\hline

 45674&F1Ia&06 28 47&$-$00 34 20&6.56&7.30&210.85&$-$05.30&$+$18.4$\pm$0.9&1.31$\pm$0.51&0.382&$\cdots$\\
180028&F6Ib&19 14 44&$+$06 02 54&6.96&7.73&040.97&$-$02.39&$-$5.1$\pm$0.7&-0.32$\pm$0.51&0.346&23.3\\
194951&F1II&20 27 07&$+$34 19 44&6.41&6.84&073.86&$-$02.34&$-$13.5$\pm$2.0&1.00$\pm$0.41&0.230&20\\
224893&A8II&00 01 37&$+$61 13 22&5.57&5.94&116.97&$-$01.07&$-$23.2$\pm$2.0&1.06$\pm$0.27&0.244&40\\

\hline
\end{tabular}}
\end{minipage}
\end{center}
\end{table*}

\section{Observations}
\label{sec:observ}

\subsection{Sample selection}
\label{sec:sample}

The stars for this analysis (HD\,45674, HD\,180028, HD\,194951 and HD\,224893)
were selected for luminous objects with classes I and II located in the galactic plane. The
derived abundances are compared with the mean abundances obtained for a sample of G--K
young supergiants of Population I taken from Luck (1977; 1978).

The stellar spectra taken for HD\,45674, HD\,180028, HD\,194951 and HD\,224893 stars come
from the library
ELODIE\footnote{http://atlas.obs-hp.fr/elodie/} originally published by Moultaka et al.\@ (2001)
and updated in its version 3.1 by Prugniel et al.\@ (2007). The spectra were observed on July
06, 1999; on July 26, 2003; on December 18, 1999 and on July 07, 1999
respectively by means of the echelle
spectrograph ELODIE placed in the 1.93m telescope located in the Haute-Provence Observatory (OHP).
They cover the spectral region between 4000 and 6800\,\AA\@,and have a resolving power
of $R$\,$\approx$\,42,000\@. The signal-to-noise ratio (S/N) for HD~45674, HD~180028,
HD\,194951 and HD\,224893 is reported to be 83, 175, 82 and 145 respectively. This S/N per
pixel has been derived for all spectra in the library at 5550\,\AA\@.

HD\,45674, HD\,180028 and HD\,194951 are objects labeled as IRAS point source, i.e.,
IRAS\,06262\,-\,0032, IRAS\,19122\,+\,0557 and IRAS\,20252\,+\,3410 respectively,
although none of them display realiable infrared excesses since
their IRAS fluxes evidence a low quality (Q=1) in 25, 60 and 100 microns. A review of the
spectral energy distributions (SED) represented from the available photometry using VisieR
database, also indicate that fluxes coming from other sources like WISE and AKARI for these three
objects does not display infrared excesses. Likewise,
HD\,224893 does not show evidence of dust around the central star.
This last object was identified by Bidelman (1993) as possible candidates to
post-AGB or normal A-and F-type supergiant stars.

Figure~\ref{fig:figure1} shows a representative range of 50\,\AA\@ of the spectrum for the stars
HD\,45674, HD\,180028, HD\,194951 and HD\,224893. The vertical dotted lines represent the position
of some absorption lines of elements such as \ion{Y}{ii}, \ion{La}{ii}, \ion{Ti}{ii},
\ion{Fe}{i} y \ion{Ni}{i} in this spectral region.
The basic parameters for the total sample are in Table~\ref{tab:table1}. This table contains the HD
number, the equatorial coordinates, the apparent $V$ and $B$ magnitude, the galactic coordinates, the
radial velocity, the parallax, the colour excess and the velocity of rotation.
These values were taken from the
SIMBAD astronomical database.\footnote{http://simbad.u-strasbg.fr/simbad/}

\begin{figure}
\centering
\includegraphics[width=7.5cm,height=7.5cm]{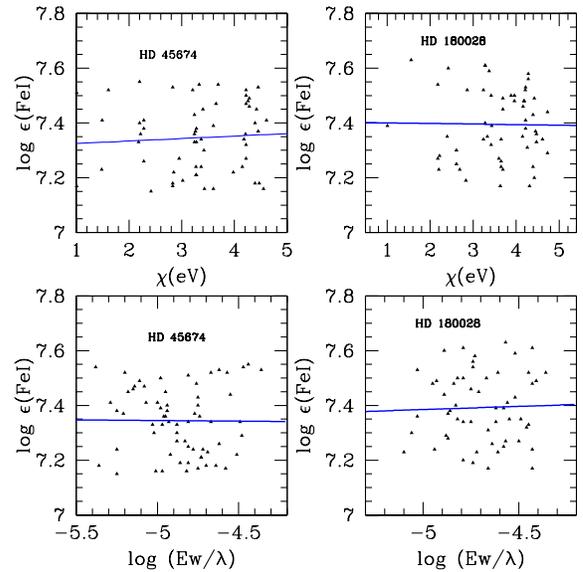}
\caption{Independence of \ion{Fe}{i} abundances from the low potential excitation and
reduced equivalent widths in HD\,45674 and HD\,180028.}
\label{fig:figure2}
\end{figure}

\begin{table*}
\begin{center}
\begin{minipage}{120mm}
  \caption{Atmospheric parameters derived in this work and those obtained from literature.}
 \label{tab:table2}
\scriptsize{\begin{tabular}{cccccr}
  \hline
  \hline
\multicolumn{1}{l}{No HD}&
\multicolumn{1}{c}{T$_\mathrm{eff}$}&
\multicolumn{1}{c}{log\,$g$}&
\multicolumn{1}{c}{$\xi_\mathrm{t}$}&
\multicolumn{1}{c}{[Fe/H]}&
\multicolumn{1}{c}{Ref.}\\
\multicolumn{1}{c}{}&
\multicolumn{1}{c}{(K)}&
\multicolumn{1}{c}{}&
\multicolumn{1}{c}{(km\,s$^{-1}$)}&
\multicolumn{1}{c}{}&
\multicolumn{1}{c}{}\\
\hline

         & 7346 & 1.95 &     &$+$0.18 & Prugniel \& Soubiran (2001)\\
45674    & 7630 & 2.05 &     &$+$0.16 & Wu et al.\@ (2011)\\
         & 7488$\pm$200 & 2.00$\pm$0.07 &     &$+$0.17$\pm$0.01 & mean value\\
         & 7500 & 2.0  & 4.0 &$+$0.16 & this work (adopted) \\
\hline
         & 6280 & 1.72 & 4.6 &$-$0.03 & Gray et al.\@ (2001b) \\
         & 6531 & 1.54 &     &$-$0.01 & Prugniel \& Soubiran (2001)\\
         & 6050 & 1.3  & 3.0 &        & Andrievsky et al.\@ (2002) \\
180028   & 6287 & 1.65 &     &$+$0.14 & Wu et al.\@ (2011)\\
         & 6307 & 1.9  & 4.0 &$+$0.10 & Kovtyukh et al.\@ (2012)\\
         & 6488 & 2.32 & 4.08&$+$0.26 & Luck (2014)\\
         & 6502 & 2.46 & 4.11&$+$0.39 & Luck (2014)\\
         & 6349$\pm$171 & 1.84$\pm$0.42 & 3.96$\pm$0.59 &$+$0.14$\pm$0.16 & mean value\\
         & 6400 & 2.1  & 4.8 &$-$0.06 & this work (adopted)\\
\hline
         & 6950 & 2.02 &     &$-$0.09 & Gray et al.\@ (2001b) \\
         & 6350 & 1.0  & 2.8 &        & Andrievsky et al.\@ (2002) \\
194951   & 7392 &      &     &        & Kovtyukh et al.\@ (2007) \\
         & 7268 &      &     &        & Hohle et al.\@ (2010) \\
         & 6760 & 1.92 &     &$-$0.13 & Lyubimkov et al.\@ (2011) \\
         & 7019 & 2.22 & 3.93&$+$0.08 & Luck (2014)\\
         & 6977 & 2.21 & 3.87&$+$0.11 & Luck (2014)\\
         & 6959$\pm$341 & 1.87$\pm$0.50 & 3.53$\pm$0.64 &$-$0.01$\pm$0.12 & mean value\\
         & 7000 & 1.8  & 4.7 &$-$0.15 & this work (adopted) \\
\hline
224893   & 7340 & 1.96 & 3.9 &$-$0.24 & Gray et al.\@ (2001b) \\
         & 7500 & 2.0  & 4.5 &$-$0.26 & this work (adopted) \\
\hline
\end{tabular}}
\end{minipage}
\end{center}
\end{table*}

\section{Atmospheric parameters}
\label{sec:param}

In order to obtain the atmospheric parameters are necessary accurate measurements of the equivalent
widths (EWs) of the Fe lines and also measurements of atomic data (log\,$g$ and $\chi$ (eV)).
The atomic data for \ion{Fe}{i} and \ion{Fe}{ii} lines were taken from F\"uhr \& Wiese (2006) and 
Mel\'endez \& Barbuy (2009).
The EWs values were restricted to range from 10 to 200\,m\AA\@, and only non-blended lines were measured.
To measure the EWs we used the task \texttt{SPLOT} of the \texttt{IRAF} software using a Gaussian fit to the
observed line profiles. The error in the measurements of EWs is about 8-10\%\@.

The stellar atmospheric models for the abundance determinations were selected from the collection done by
Castelli and Kurucz (2003). These models have been constructed with a plane-parallel geometry,
hydrostatic equilibrium, local thermodynamic equilibrium (LTE) and the \texttt{ODFNEW} opacity distribution.

We used the updated version 2010 of the \texttt{MOOG} code (Sneden 1973) in the determination of the atmospheric
parameters. This code has been developed under ETL assumptions. Based on trial and error test, we selected
the model that best fits the photosphere of stars. With the best model, the effective temperature was
determined under the assumption of excitation equilibrium, i.e., by
requiring that the derived iron abundances should be independent of the low excitation potential.
The surface gravity is derived
by requesting that the \ion{Fe}{i} abundances be similar to those obtained for the \ion{Fe}{ii}. 
To confirm the value of
gravity, the equilibrium condition can be extended to other elements with two ionization states
like Ca, Ti, Cr, Ni. Finally the microturbulence velocity is determined by imposing that the abundances of
\ion{Fe}{i} are independent of reduced equivalent widths (log\,Ew/$\lambda$).
Figure~\ref{fig:figure2} shows the independence between the abundances of \ion{Fe}{i} concerning to the low
potential excitation and reduced equivalent widths for HD\,45674 and HD\,180028.

From Table~\ref{tab:table2} can be identified the atmospheric parameters estimated by
different authors from photometric and spectroscopic techniques and also the atmospheric parameters
adopted from this work. For each object with more than one measures it was estimated a mean
value in their atmospheric parameters. We note that our adopted atmospheric parameters match satisfactorily
with these mean values.

\begin{table*}
\begin{minipage}{180mm}
\caption{Sensitivity of abundances to the incertainties in the model parameters for two range of temperature
covering our sample stars.}
\label{tab:table3}
\begin{center}
\scriptsize{\begin{tabular}{lcrrrcrrcrc}
\noalign{\smallskip}
\hline
\hline
\noalign{\smallskip}
\noalign{\smallskip}
    &\multicolumn{5}{c}{HD\,45674} &\multicolumn{5}{c}{HD\,180028} \\
    &\multicolumn{5}{c}{(7500\,K)} &\multicolumn{5}{c}{(6400\,K)}   \\
\multicolumn{1}{l}{Species}&
\multicolumn{1}{c}{$\Delta$T$_\mathrm{eff}$}&
\multicolumn{1}{c}{$\Delta$log\,$g$}&
\multicolumn{1}{c}{$\Delta$$\xi_\mathrm{t}$}&
\multicolumn{1}{c}{$\Delta$Ew}&
\multicolumn{1}{c}{$\sigma_\mathrm{tot}$}&
\multicolumn{1}{c}{$\Delta$T$_\mathrm{eff}$}&
\multicolumn{1}{c}{$\Delta$log\,$g$}&
\multicolumn{1}{c}{$\Delta$$\xi_\mathrm{t}$}&
\multicolumn{1}{c}{$\Delta$Ew}&
\multicolumn{1}{c}{$\sigma_\mathrm{tot}$}\\
\multicolumn{1}{c}{}&
\multicolumn{1}{c}{$+$250\,K}&
\multicolumn{1}{c}{$+$0.5}&
\multicolumn{1}{c}{$+$0.5}&
\multicolumn{1}{c}{$+$10\%}&
\multicolumn{1}{c}{}&
\multicolumn{1}{c}{$+$250\,K}&
\multicolumn{1}{c}{$+$0.5}&
\multicolumn{1}{c}{$+$0.5}&
\multicolumn{1}{c}{$+$10\%}&
\multicolumn{1}{c}{}\\
           \noalign{\smallskip}
            \hline
            \noalign{\smallskip}
\ion{C}{i}& $-0.14$ & $ 0.00$ & $+0.01$ & $-0.01$ & $0.14$ & $+0.07$  & $-0.11$  & $-0.01$  & $ 0.00$  & $0.13$\\
\ion{N}{i}& $-0.02$ & $-0.06$ & $ 0.00$ & $-0.01$ & $0.06$ & $\cdots$ & $\cdots$ & $\cdots$ & $\cdots$ &$\cdots$\\
\ion{O}{i}& $+0.02$ & $-0.08$ & $+0.01$ & $-0.01$ & $0.08$ & $+0.10$  & $-0.05$  & $ 0.00$  & $\cdots$ & $0.11$\\
\ion{Na}{i}& $-0.20$ & $+0.08$ & $+0.01$ & $-0.01$ & $0.22$ & $-0.03$  & $-0.03$  & $+0.05$  & $-0.01$  & $0.07$\\
\ion{Mg}{i}& $-0.22$ & $+0.08$ & $+0.08$ & $-0.01$ & $0.25$ & $-0.14$  & $-0.04$  & $+0.04$  & $ 0.00$  & $0.15$\\
\ion{Si}{i}& $\cdots$& $\cdots$& $\cdots$& $\cdots$& $\cdots$&$-0.03$  & $-0.03$  & $+0.01$  & $ 0.00$  & $0.04$\\
\ion{Si}{ii}& $+0.02$ & $-0.08$ & $+0.16$ & $-0.02$ & $0.18$ & $+0.13$  & $-0.15$  & $+0.07$  & $-0.01$  & $0.21$\\
\ion{S}{i}& $-0.17$ & $+0.04$ & $+0.02$ & $-0.01$ & $0.18$ & $+0.04$  & $-0.11$  & $+0.02$  & $-0.01$  & $0.12$\\
\ion{Ca}{i}& $-0.27$ & $+0.10$ & $+0.05$ & $-0.01$ & $0.29$ & $-0.07$  & $-0.02$  & $+0.04$  & $-0.01$  & $0.08$\\
\ion{Ca}{ii}& $-0.07$ & $-0.04$ & $+0.03$ & $ 0.00$ & $0.09$ & $\cdots$ & $\cdots$ & $\cdots$ & $\cdots$ &$\cdots$\\
\ion{Sc}{ii}& $-0.15$ & $-0.12$ & $+0.07$ & $-0.01$ & $0.20$ & $-0.04$  & $-0.16$  & $+0.03$  & $-0.01$  & $0.17$\\
\ion{Ti}{i}& $-0.28$ & $+0.05$ & $ 0.00$ & $-0.02$ & $0.29$ & $-0.14$  & $-0.04$  & $+0.01$  & $-0.01$  & $0.15$\\
\ion{Ti}{ii}& $-0.15$ & $-0.13$ & $+0.08$ & $-0.02$ & $0.21$ & $-0.03$  & $-0.16$  & $+0.04$  & $-0.01$  & $0.17$\\
\ion{V}{ii}& $-0.13$ & $-0.12$ & $+0.02$ & $-0.01$ & $0.18$ & $\cdots$ & $\cdots$ & $\cdots$ & $\cdots$ &$\cdots$\\
\ion{Cr}{i}& $-0.27$ & $+0.06$ & $+0.03$ & $-0.01$ & $0.28$ & $-0.12$  & $-0.04$  & $+0.03$  & $-0.01$  & $0.13$\\
\ion{Cr}{ii}& $-0.10$ & $-0.12$ & $+0.05$ & $-0.01$ & $0.16$ & $+0.03$  & $-0.16$  & $+0.06$  & $-0.02$  & $0.17$\\
\ion{Mn}{i}& $-0.24$ & $+0.07$ & $+0.02$ & $-0.01$ & $0.25$ & $-0.08$  & $-0.04$  & $+0.04$  & $-0.01$  & $0.10$\\
\ion{Fe}{i}& $-0.24$ & $+0.07$ & $+0.03$ & $-0.01$ & $0.25$ & $-0.08$  & $-0.04$  & $+0.04$  & $-0.01$  & $0.10$\\
\ion{Fe}{ii}& $-0.10$ & $-0.12$ & $+0.04$ & $-0.01$ & $0.16$ & $+0.02$  & $-0.15$  & $+0.05$  & $-0.01$  & $0.16$\\
\ion{Ni}{i}& $-0.22$ & $+0.08$ & $+0.01$ & $ 0.00$ & $0.23$ & $-0.08$  & $-0.04$  & $+0.02$  & $-0.01$  & $0.09$\\
\ion{Cu}{i}& $\cdots$& $\cdots$& $\cdots$& $\cdots$&$\cdots$& $-0.13$  & $-0.04$  & $+0.01$  & $-0.01$  & $0.14$\\
\ion{Zn}{i}& $-0.24$ & $+0.07$ & $+0.01$ & $ 0.00$ & $0.25$ & $-0.08$  & $-0.05$  & $+0.02$  & $-0.01$  & $0.10$\\
\ion{Y}{ii}& $-0.19$ & $-0.11$ & $+0.03$ & $-0.01$ & $0.22$ & $-0.04$  & $-0.16$  & $+0.02$  & $ 0.00$  & $0.17$\\
\ion{Zr}{ii}& $-0.17$ & $-0.13$ & $+0.05$ & $-0.02$ & $0.22$ & $-0.03$  & $-0.16$  & $+0.01$  & $ 0.00$  & $0.16$\\
\ion{Ba}{ii}& $-0.30$ & $ 0.00$ & $+0.05$ & $-0.01$ & $0.30$ & $-0.10$  & $-0.10$  & $+0.07$  & $\cdots$ & $0.15$\\
\ion{La}{ii}& $\cdots$& $\cdots$& $\cdots$& $\cdots$&$\cdots$& $-0.08$  & $-0.16$  & $+0.06$  & $-0.02$  & $0.19$\\
\ion{Ce}{ii}& $-0.24$ & $-0.08$ & $+0.01$ & $-0.01$ & $0.25$ & $-0.08$  & $-0.15$  & $+0.01$  & $ 0.00$  & $0.17$\\
\ion{Nd}{ii}& $-0.28$ & $-0.05$ & $ 0.00$ & $ 0.00$ & $0.28$ & $-0.11$ & $-0.15$   & $+0.01$  & $ 0.00$  & $0.19$\\
\ion{Eu}{ii}& $\cdots$& $\cdots$& $\cdots$& $\cdots$&$\cdots$& $-0.10$  & $-0.18$  & $ 0.00$  & $\cdots$ & $0.20$\\
            \noalign{\smallskip}
            \hline
            \noalign{\smallskip}
\end{tabular}}
\end{center}
\end{minipage}
\end{table*}

\subsection{Uncertainty in the abundances}
\label{sec:uncertain}

We calculate the effects on chemical abundances due to errors in the measured equivalent widths
(8--10\%\@), the defined model of atmospheric parameters ($+$250\,K in T$_\mathrm{eff}$, $+$0.5 in log\,$g$
and $+$0.5 km\,s$^{-1}$ in $\xi_\mathrm{t}$), and the atomic data (log\,$gf$ and $\chi$ (eV)).
Errors due to equivalent widths are random because they depend on several factors such as the
position of the continuum, the signal to noise ratio (S/N) and the spectral type of the star.

In contrast errors due to the atmospheric parameters and atomic data are systematic and depend
on the quality with which they were derived.
In short, the error in the $gf$-values can varies from element to element.
For example, experimental values of high accuracy for \ion{Fe}{i} and \ion{Fe}{ii} between
3\%\@ to 10\%\@ are available for a large fraction of lines.
For other Fe-peak elements, errors in their $gf$-values may range from 10 to 25\%\@. For neutron-
capture elements the accuracy is the 10 to 50\%\@ range.
The sensitivity of derived abundances to changes in the model atmosphere parameters are
described in Table~\ref{tab:table3}.

HD\,45674 and HD\,194951 have the same temperature, i.e., 7500\,K.
For these stars the spectra are much cleaner and the temperature are not large enough for the lines
to develop strong wings making line strengths inaccurate. We can see that a variation in effective
temperature of 250\,K generates greater uncertainty in all lines, particularly in neutral lines
except C, N, O and ionized lines like Ba, Ce and Nd. Contrary
we did not observe significant changes in the uncertainties due to changes in log\,$g$ by 0.5,
$\xi_\mathrm{t}$ by 0.5 km\,s$^{-1}$ and Ew by 10 per cent.
On the other hand, a lower temperature (6400\,K) the line strenghts
shows sensitivity only to changes either in T$_\mathrm{eff}$ by 250\,K and log\,$g$ by 0.5.

The total error $\sigma_\mathrm{tot}$ for each element is given by the square root of the quadratic
sum of the random and systematic errors.

\section{Determination of the abundances}
\label{sec:abund}

In the determination of chemical abundances we use the equivalent widths of 176, 116,
184 and 73 absorption lines identified in HD\,45674, HD\,180028, HD\,194951
and HD\,224893 respectively.
For this purpose we use the task \texttt{ABFIND} of the \texttt{MOOG} code (Sneden 1973).
We employ the atmospheric models adopted in Table~\ref{tab:table2}.
The sources of the $gf$-values for different elements are those that have been
referred by Sumangala Rao, Giridhar and Lambert (2012) (see Table 4 for further
details).

Sumangala Rao, Giridhar and Lambert (2012)
also studied the systematic differences caused when they made use different sources for $gf$-values.
The authors employed for this purpose the solar spectrum taken from Solar Flux Atlas (Kurucz et al. 1984)
and found very few differences to solar values for most of the elements, i.e., between 0.04 to 0.09 dex.

The chemical abundances derived for HD\,45674, HD\,180028, HD\,194951 and HD\,224893 can be seen in
Table~\ref{tab:table4}.
This table contains the chemical species present in the photosphere, the solar photospheric abundances
are given
by Asplund et al. (2005), as well as those calculated for each stars and its uncertainty,
the abundances relative to hydrogen, the
number of identified lines and the abundances relative to iron.
The abundances of the elements in Table~\ref{tab:table4} are in a logarithmic scale with respect
to hydrogen, namely: log $\epsilon$(X) = log [N(X)/N(H)] + 12.0. The abundances of the elements
relative to hydrogen and iron are expressed as [X/H] = log $\epsilon$(X)star -
log $\epsilon$(X)sun and [X/Fe] = [X/H] - [Fe/H] respectively.

\begin{table*}
\begin{minipage}{170mm}
\caption{Elemental abundances for HD\,45674, HD\,180028, HD\,194951 and HD\,224893.}
\label{tab:table4}
\begin{center}
\tiny{\begin{tabular}{lcrrrccrccrccr}
\noalign{\smallskip}
\hline
\hline
\noalign{\smallskip}
\noalign{\smallskip}
   &   &\multicolumn{3}{c}{HD\,45674} &\multicolumn{3}{c}{HD\,180028} &\multicolumn{3}{c}{HD\,194951} &
        \multicolumn{3}{c}{HD\,224893}\\
\cline{3-14} \\
\multicolumn{1}{l}{Species}&
\multicolumn{1}{c}{$\log \epsilon_{\odot}$}&
\multicolumn{1}{c}{[X/H]}&
\multicolumn{1}{c}{N}&
\multicolumn{1}{r}{[X/Fe]}&
\multicolumn{1}{c}{[X/H]}&
\multicolumn{1}{c}{N}&
\multicolumn{1}{r}{[X/Fe]}&
\multicolumn{1}{c}{[X/H]}&
\multicolumn{1}{c}{N}&
\multicolumn{1}{r}{[X/Fe]}&
\multicolumn{1}{c}{[X/H]}&
\multicolumn{1}{c}{N}&
\multicolumn{1}{r}{[X/Fe]}\\
           \noalign{\smallskip}
            \hline
            \noalign{\smallskip}
\ion{C}{i}& 8.39 & $-0.20$$\pm$0.02& 2 &$-0.36$ & $-0.28$$\pm$0.10&syn&$-0.22$ & $-0.16$$\pm$0.09& 5 &$-0.01$
            & $-0.11$$\pm$0.10& 2 &$+0.15$\\
\ion{N}{i}& 7.78 & $+0.68$$\pm$0.06& 1 &$+0.52$ &                 &   &        &                 &   &
            & $+0.89$$\pm$0.06& 1 &$+1.15$\\
\ion{O}{i}& 8.66 & $+0.03$$\pm$0.01& 2 &$-0.13$ & $+0.07$$\pm$0.10&syn&$+0.13$ & $-0.10$$\pm$0.00& 2 &$+0.05$
            & $+0.19$$\pm$0.05& 1 &$+0.45$\\
\ion{Na}{i}& 6.17 & $+0.50$$\pm$0.10& 4 &$+0.34$ & $+0.37$$\pm$0.04& 2 &$+0.43$ & $+0.09$$\pm$0.00& 2 &$+0.24$
            & $+0.29$$\pm$0.04& 1 &$+0.55$\\
\ion{Mg}{i}& 7.53 & $+0.08$$\pm$0.19& 4 &$-0.08$ & $+0.19$$\pm$0.09& 1 &$+0.25$ & $-0.21$$\pm$0.09& 5 &$-0.06$
            & $-0.34$$\pm$0.11& 3 &$-0.08$\\
\ion{Si}{i}& 7.51 &                 &   &        & $+0.19$$\pm$0.14& 3 &$+0.25$ & $+0.18$$\pm$0.06& 2 &$+0.33$
            &                 &   &       \\
\ion{Si}{ii}& 7.51 & $+0.47$$\pm$0.13& 2 &$+0.31$ & $+0.06$$\pm$0.04& 1 &$+0.12$ &                 &   &
            & $-0.28$$\pm$0.29& 2 &$-0.02$\\
\ion{S}{i}& 7.14 & $+0.41$$\pm$0.10& 5 &$+0.25$ & $+0.38$$\pm$0.09& 3 &$+0.44$ & $-0.03$$\pm$0.10&syn&$+0.12$
            &                 &   &       \\
\ion{Ca}{i}& 6.31 & $+0.24$$\pm$0.11& 8 &$+0.08$ & $+0.07$$\pm$0.13& 5 &$+0.13$ & $-0.13$$\pm$0.17& 10&$+0.02$
            & $-0.29$$\pm$0.03& 4 &$-0.03$\\
\ion{Ca}{ii}& 6.31 & $+0.16$$\pm$0.15& 2 &$ 0.00$ &                 &   &        &                 &   &
            &                 &   &       \\
\ion{Sc}{ii}& 3.05 & $+0.34$$\pm$0.04& 9 &$+0.18$ & $+0.25$$\pm$0.09& 4 &$+0.31$ & $-0.03$$\pm$0.14& 6 &$+0.12$
            & $+0.13$$\pm$0.05& 2 &$+0.39$\\
\ion{Ti}{i}& 4.90 & $+0.27$$\pm$0.10& 3 &$+0.11$ & $+0.15$$\pm$0.06& 1 &$+0.21$ & $-0.21$$\pm$0.06& 2 &$-0.06$
            &                 &   &       \\
\ion{Ti}{ii}& 4.90 & $+0.29$$\pm$0.14&14 &$+0.13$ & $+0.05$$\pm$0.10& 4 &$+0.11$ & $-0.18$$\pm$0.12&13 &$-0.03$
            & $-0.20$$\pm$0.11&13 &$+0.06$\\
\ion{V}{ii}& 4.00 & $+0.22$$\pm$0.02& 1 &$+0.06$ &                 &   &        &                 &   &
            & $-0.24$$\pm$0.02& 1 &$+0.02$ \\
\ion{Cr}{i}& 5.64 & $+0.13$$\pm$0.08& 3 &$-0.03$ & $-0.11$$\pm$0.10& 2 &$-0.05$ & $-0.34$$\pm$0.13& 5 &$-0.19$
            & $-0.25$$\pm$0.12& 2 &$+0.01$\\
\ion{Cr}{ii}& 5.64 & $+0.18$$\pm$0.12& 7 &$+0.02$ & $-0.11$$\pm$0.13& 3 &$-0.07$ & $-0.25$$\pm$0.11& 11&$-0.10$
            & $-0.40$$\pm$0.09& 4 &$+0.14$\\
\ion{Mn}{i}& 5.39 & $+0.14$$\pm$0.01& 3 &$-0.02$ & $+0.11$$\pm$0.08& 4 &$+0.17$ & $-0.21$$\pm$0.18& 5 &$-0.06$
            & $-0.21$$\pm$0.03& 1 &$+0.05$\\
\ion{Fe}{i}& 7.45 & $+0.12$$\pm$0.10&60 &        & $-0.06$$\pm$0.13&56 &        & $-0.17$$\pm$0.11& 72&
            & $-0.27$$\pm$0.11&19 &       \\
\ion{Fe}{ii}& 7.45 & $+0.20$$\pm$0.08&10 &        & $-0.06$$\pm$0.16& 8 &        & $-0.12$$\pm$0.11& 20&
            & $-0.26$$\pm$0.17& 7 &       \\
\ion{Ni}{i}& 6.23 & $-0.06$$\pm$0.12& 8 &$-0.22$ & $-0.09$$\pm$0.06& 4 &$-0.03$ & $+0.10$$\pm$0.11& 6 &$+0.25$
            &                 &   &       \\
\ion{Ni}{ii}& 6.23 &                 &   &        &                 &   &        & $+0.01$$\pm$0.04& 1 &$+0.16$
            &                 &   &       \\
\ion{Cu}{i}& 4.21 &                 &   &        & $-0.26$$\pm$0.04& 1 &$-0.20$ &                 &   &
            &                 &   &       \\
\ion{Zn}{i}& 4.60 & $-0.17$$\pm$0.16& 2 &$-0.33$ & $-0.38$$\pm$0.03& 1 &$-0.32$ & $-0.20$$\pm$0.10&syn&$-0.05$
            & $-0.26$$\pm$0.03 & 1 &$ 0.00$\\
\ion{Y}{ii}& 2.21 & $+0.11$$\pm$0.10& 5 &$-0.05$ & $+0.05$$\pm$0.12& 3 &$+0.11$ & $-0.20$$\pm$0.15& 5 &$-0.05$
            & $-0.20$$\pm$0.09& 4 &$+0.06$\\
\ion{Zr}{ii}& 2.59 & $+0.61$$\pm$0.13& 3 &$+0.45$ & $+0.10$$\pm$0.04& 1 &$+0.16$ & $-0.12$$\pm$0.19& 2 &$+0.03$
            &                 &   &       \\
\ion{Ba}{ii}& 2.17 & $+0.35$$\pm$0.07& 1 &$+0.19$ & $+0.23$$\pm$0.10&syn&$+0.29$ & $-0.11$$\pm$0.07& 1 &$+0.04$
            & $-0.34$$\pm$0.10& 2 &$-0.08$\\
\ion{La}{ii}& 1.13 &                 &   &        & $+0.29$$\pm$0.05& 1 &$+0.35$ &                 &   &
            &                 &   &        \\
\ion{Ce}{ii}& 1.58 & $+0.33$$\pm$0.14& 3 &$+0.17$ & $-0.08$$\pm$0.16& 3 &$-0.02$ &                 &   &
            &                 &   &        \\
\ion{Nd}{ii}& 1.45 & $+0.22$$\pm$0.05& 1 &$+0.06$ & $+0.13$$\pm$0.09& 2 &$+0.19$ & $-0.11$$\pm$0.17& 2 &$+0.04$
            &                 &   &        \\
\ion{Eu}{ii}& 0.52 &                 &   &        & $-0.12$$\pm$0.10&syn&$-0.06$ &                 &   &
            &                 &   &      \\
            \noalign{\smallskip}
            \hline
            \noalign{\smallskip}
\end{tabular}}
\end{center}
\end{minipage}
\end{table*}

\section{Individual discussion of the abundances}
\label{sec:discuss}

\subsection{HD\,45674}
\label{sec:hd45674}

The star HD\,45674 has been classified as F1Ia.
Bidelman (1993) includes this object as a possible candidate to
post-AGB or A-and-F normal supergiant stars.
Their atmospheric parameters were obtained from ionization
equilibrium between the lines of \ion{Fe}{i} and \ion{Fe}{ii} reaching values of 
T$_\mathrm{eff}$\,=\,7500\,K,
log\,$g$\,=\,2.0, $\chi_\mathrm{t}$\,=\,4.0 and [Fe/H]\,=\,$+$0.16 dex respectively.
The surface gravity for HD\,45674 was re-affirmed
from the equilibrium ionization of other elements such as Ca, Ti, Cr, namely
$\Delta$\,=\,[X$_\mathrm{II}$/H]--[X$_\mathrm{I}$/H] is 0.08, $-$0.02 and $-$0.05 respectively.
A recopilation of the atmospheric parameters obtained by different authors are shown
in Table~\ref{tab:table2}.

With respect to the abundances of the light elements CNO, the carbon abundance is obtained
from two lines (4769\,\AA\@ and 5380\,\AA\@) being their values [C/H]\,=\,$-$0.20 and [C/Fe]\,=\,$-$0.36.
Venn (1995b) predicts a non-LTE correction of $-$0.25 in the C abundance for the supergiant
HD\,36673 (F0Ib) with effective temperature of 7400\,K.
HD\,45674 has an effective temperature of 7500\,K therefore none non-LTE correction must
be made. Taking a value of $-$0.25 dex leads to an abundance of [C/Fe]\,=\,$-$0.69.

The
N abundance for HD\,45674 shows a moderate enrichement ([N/Fe]\,=\,$+$0.68), in fact this abundance
is obtained from $\lambda$4151.4\,\AA\@ line with a EW of 10.1\,m\AA\@.
A non-LTE correction of $-$0.28 is derived by Venn (1995b) in the near IR region at 7400\,K for
HD\,196379 with A9II type. The non-LTE correction made for \ion{N}{i} by Luck \& Lambert (1985)
indicates that at the temperature of 7500\,K, an equivalent width of $\sim$10\,m\AA\@ and log\,$g$\,=\,
2.0 its value is almost zero (see Figure 8).
By contrast, for the blue region Przybilla \& Butler (2001) found a non-LTE
correction of $-$0.11 dex for \ion{N}{i} ($\lambda$3830\,\AA\@) at a temperature of 9600\,K and
where its value increases negatively toward the IR lines. Taking into consideration the above we can
argue that the non-LTE correction should be small for this line.

We adopted a value of $\sim$$-$0.08 dex for our line in the
blue region. Our non-LTE values are [N/H]\,=\,$+$0.60 dex and [N/Fe]\,=\,$+$0.36 dex
respectively. The observed deficiency in C and the moderate enhancement in N involves
a conversion of initial
carbon into N through CN-cycle and products of the 1DUP are brought to the surface.

The O abundance is near to solar value
([O/Fe]\,=\,$+$0.03) and is obtained from $\lambda$6156.0\,\AA\@ and $\lambda$6156.8\,\AA\@ lines.
To this temperature a non-LTE correction about of $-$0.15 taken from Takeda \& Takada-Hidai (1998)
would lead to abundances of [O/H]\,=\,$-$0.12 and [O/Fe]\,=\,$-$0.36 respectively.
The C/O ratio of 0.32 indicates that HD\,45674 is O-rich.

Iron-peak elements (Sc, V, Cr, Ni) shows [X/Fe] abundances nearest to solar and ranges
from $-$0.22 to $+$0.18 dex.
The [$\alpha$/Fe] ratio $\sim$$+$0.10$\pm$0.16 estimated from Mg, Si, Ca, Ti shows a typical value
for objects of thin disk (see Reddy et al.\@ 2003; 2006). Si and S abundances
are moderate enhancements, i.e.,
[Si/Fe]\,=\,$+$0.31 and [S/Fe]\,=\,$+$0.25 respectively.
The Zn abundance is derived from two lines at $\lambda$4722\,\AA\@ and $\lambda$4810\,\AA\@. The value
of log\,$gf$\,=\,$-$0.25 for the line at $\lambda$4810\,\AA\@ has been obtained from Barbuy et at. (2015).
[Zn/Fe]\,=\,$-$0.33 is lower than expected for objects of the thin and thick
disk population, i.e., $\sim$ $-$0.20 a $-$0.30 dex (Reddy et al.\@ 2003; 2006).
Neutron-capture elements shows [X/Fe] abundances with tendency towards to solar value and
ranges from $+$0.06 to $+$0.19 dex. Of the elements identified like \ion{Y}{ii}, \ion{Ba}{ii}, \ion{Ce}{ii} 
and \ion{Nd}{ii}
only the \ion{Zr}{ii} seems to be found moderately enriched ([Zr/Fe]\,=\,$+$0.45).

\subsection{HD\,180028}
\label{sec:hd1800}

HD\,180028 is classified as a F6Ib and cataloged as \texttt{IRAS} point source (IRAS~19122\,+\,0557). The
radial velocity were reported by different authors; $-$6.0\,km\,$^{-1}$ (Wilson 1953),
$-$6.0\,km\,$^{-1}$ (Duflot et al.\@ 1995) and $-$5.1\,km\,$^{-1}$ (Gontcharov 2006) and does not
present variability. Their absorption lines appear slightly wider indicating that rotates with velocity
of 23.3\,km\,$^{-1}$ (De Medeiros et al.\@ 2002).

Their atmospheric parameters were obtained from ionization
equilibrium between the lines of \ion{Fe}{i} and \ion{Fe}{ii} reaching values of T$_\mathrm{eff}$\,=\,6400\,K,
log\,$g$\,=\,2.1, $\chi_\mathrm{t}$\,=\,4.8 and [Fe/H]\,=\,$-$0.06 dex respectively.
The surface gravity for HD\,180028 was re-affirmed
from the equilibrium ionization of other elements such as Si, Ti, Cr, namely
$\Delta$\,=\,[X$_\mathrm{II}$/H]--[X$_\mathrm{I}$/H] is 0.11, $-$0.10 and $-$0.00 respectively.
The atmospheric parameters obtained by different
authors and the adopted parameters are found in Table~\ref{tab:table2}.

\begin{figure}
\centering
\includegraphics[width=7.5cm,height=7.5cm]{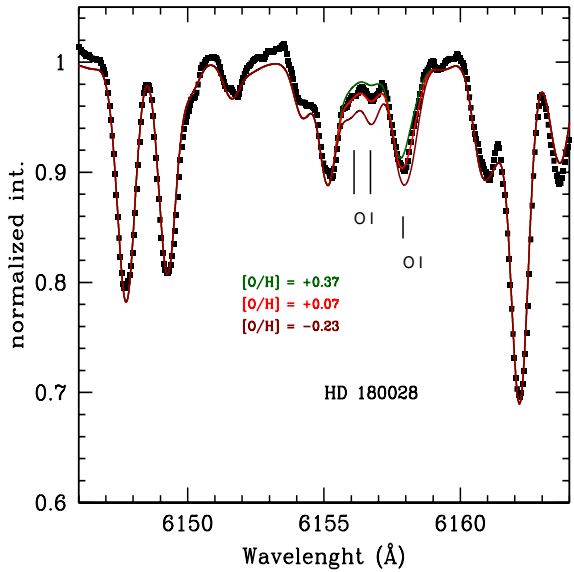}
\caption{Observed (filled squares) and synthetic (solid line) spectra of the \ion{O}{i} triplet at
6140--60\,\AA\@ region. The synthetic spectra correspond from top to bottom to [O/H]\,=\,$=$0.37,
$=$0.07 and $-$0.23 dex respectively.}
\label{fig:figure3}
\end{figure}

HD\,180028 exhibits a metallicity near to solar of [Fe/H] of$-$0.06.
The abundances of C was obtained by synthesizing the line at 5380\,\AA\@ reaching values of
[C/H]\,=\,$-$0.28 dex
and [C/Fe]\,=\,$-$0.22 dex respectively. At temperature of 6100\,K the non-LTE correction to C is
very small
(Takeda 1994) and leads to a value of [C/Fe]\,=\,$-$0.39 dex. In their spectrum are not present N lines
or CN bands employee for estimate its abundances.
The O abundance was derived by synthesizing the region at 6140--60\,\AA\@ and where their values 
are [O/H]\,=\,$+$0.07 dex and [O/Fe]\,=\,$+$0.13 dex respectively (see Figure~\ref{fig:figure3}.
These values are higher than obtained by Luck (2014). The C/O ratio is about of $\sim$0.23 which
is lower than solar value.
From sodium analysis conducted by Andrievsky et al.\@ (2002) found a moderate enrichment of
[Na/Fe]\,=\,$+$0.28 dex. The value obtained in this work is slightly higher, i.e.
[Na/Fe]\,=\,$+$0.43 dex. The Fe-peak elements show a solar trend. The ratio [$\alpha$/Fe] of $+$0.18
$\pm$0.05 dex seems to indicate that this object belongs to thin disk. Individually, Si, Ca and Ti present
solar values while Mg shows a moderate enhancement [Mg/Fe]\,=\,$+$0.25 dex. Sulfur is found be enriched,
i.e., [S/Fe]\,=\,$+$0.44 dex. The s-elements show solar-trend except Ba and La which show a modest
enhancement.

\subsection{HD\,194951}
\label{sec:hd1949}

HD\,194951 was classified with F-type by different authors; F2Iab by Stock et al.\@(1960),
F3II by Bidelman (1957) and F1II by Morgan (1972).
The physical parameters have been determined by various authors from photometric and
spectroscopic techniques (see Table~\ref{tab:table2}) while the values adopted can be seen
in the Table~\ref{tab:table3}. The superficial gravity of HD\,194951 was re-affirmed
from the equilibrium ionization of other elements such as Ti, Cr, and Ni, that is
$\Delta$\,=\,[X$_\mathrm{II}$/H]--[X$_\mathrm{I}$/H] is $+$0.05, $+$0.09 and $-$0.09 respectively.
The chemical abundances of the elements identified are represented in Table~\ref{tab:table4}.

Carbon abundance is determined from five lines ($\lambda$4769.9\,\AA\@, $\lambda$4775.8\,\AA\@,
$\lambda$4932.0\,\AA\@, $\lambda$5380.3\,\AA\@ and $\lambda$6014.8\,\AA\@) reaching a [C/H] value
of $-$0.16. At a temperature of 7000\,K the non-ETL effects are present and a correction
of $-$0.2 dex for carbon leads to values of [C/H]\,=\,$-$0.36 and [C/Fe]\,=\,$-$0.26
respectively. For O abundance a non-LTE correction of $-$0.1 dex taken from Takeda \&
Takada-Hidai (1998) is estimated and the abundances [O/H]\,=\,$-$0.20 and [O/Fe]\,=\,
$-$0.10 are
near to solar. These C and O abundance values are similar to obtained by Luck (2014).
The N line at $\lambda$4151.4\,\AA\@ found for HD\,45674 and HD\,224893 could not be measured
for this spectrum.
However, Lyubimkov et al.\@ (2011) were able to study the N abundance
for HD\,194951 using high-resolution spectra and they found that N is enriched with values of
[N/H]\,=\,$+$0.60 and [N/Fe]\,=\,$+$0.73$\pm$0.18 respectively.

Our atmospheric model differs
in 240\,K and 0.12\@ in effective temperature and surface gravity relative to the model used by
Lyumbikov et al.\@, i.e., 6760\,K and 1.92\@. In Table~\ref{tab:table3} we observe that very
few changes in the N abundance appears due to variations of 250\,K and 0.5\@ in T$_\mathrm{eff}$ and
log\,$g$. This means we can use this N abundance without being affected by the model.
According to the authors the N enrichment has been a
general characteristic present in supergiant stars with A and F type that have experienced
the first dredge-up (1DUP). The deficiency in C and the enrichment of N indicates that has
operated the CN-cycle and the processed material from the H-burning is released on the
surface of the star. The C/O\,=\,0.46 ratio is practically solar and it is O-rich star.

The [$\alpha$/Fe] ratio
of $+$0.06 presents a consistent value for an object belonging to the thin disk population.
The Si abundance exhibit a moderate enrichment, i.e., [Si/Fe]\,=\,$+$0.33.
Other elements such as Sc, Cr, Mn and Zn shows abundances [X/Fe] between $-$0.15 up to $+$0.09
similar to those presented in objects of the thin and thick disk. The Zn abundance is obtained
from synthesized line at $\lambda$4810.5\,\AA\@.
[X/Fe] of s-elements range from $-$0.09 to $+$0.01, that is, they are on average solar.

\subsection{HD\,224893}
\label{sec:hd2248}

Gray et al.\@ (2001b) have determined their atmospheric parameters for this star (see Table~\ref
{tab:table2})
and was classified as a bright giant with A8II. Several authors have studied the radial velocity
in different epochs, i.e., $-$22.4 by Adams (1915), $-$26.4 by Harper (1937), $-$23.2 by Wilson
(1953), $-$21.9 by Bouigue et al.\@ (1953), $-$27 by Fehrenbach et al.\@ (1996)
and $-$25.10\,km\,s$^{-1}$ by Gontcharov (2006)
respectively, showing very little variations. Danziger \& Faber (1972) report a rotation velocity
of 40\,km\,s$^{-1}$.

In this paper is performed for the first time an analysis of abundances in which is included a total
of 15 elements. The abundances can be seen in the Table~\ref{tab:table4}.
The metallicity of HD\,224893 is moderately deficient ([Fe/H]\,=\,$-$0.26 dex).
The C abundance is obtained from lines 4770.0\,\AA\@ and 4771.7\,\AA\@ respectively. From the LTE
analysis we derive a value of [C/H] of $-$0.11 dex. By taking a non-LTE correction of $-$0.26 dex
(Venn 1995b), either [C/H] and [C/Fe] reaches values of $-$0.37 dex and $-$0.19 dex respectively.
The N abundance obtained from $\lambda$4151.4\,\AA\@ are [N/H]\,=\,$+$0.89 dex and [N/Fe]\,=\,$+$1.15
dex respectively. The equivalent width of this line is 15.8\,m\AA\@.
Under a non-LTE correction of $-$0.08 dex adopted at 7500\@K
their new values are [N/H]\,=\,$+$0.81 dex and [N/Fe]\,=\,$+$0.99 dex respectively. Like
HD~45674 this star shows evidence of having operated CN-cycle and the 1DUP event.

The O abundance is derived from the line at 6158\,\AA\@ and shows sign of overabundance ([O/Fe]\,
=\,$+$0.45), however a non-LTE correction of $-$0.15 dex at temperature of 7500\,K taking from
Takeda \& Takada-Hidai (1998) leads to values of [O/H]\,=\,$+$0.35 dex and [O/Fe]\,=\,$+$0.53
dex respectively. These values indicate that O abundance show a moderate enrichment.
The ratio C/O\,=\,0.26 indicate that HD\,224893 is O-rich and has a similar value to HD\,45674.

The abundances [X/Fe] of Fe-peak elements as V, Cr and Mn show solar values except the Sc
which has a moderate enrichmenet of $+$0.39 dex. The alpha-elements ([$\alpha$/Fe]\,=\,$-$0.02 dex),
the Zn and the s-elements (Y and Ba) shown sunlike tendencies.

\section {Discussion}
\label{sec:results}

\subsection{Atmospheric parameters}
\label{sec:atmosparameters}

The atmospheric parameters taken from literature and those adopted in this work are present in
Table~\ref{tab:table2}. We take a mean value of atmospheric parameters of those objects
with more than two values. We can note that our adopted values of effective temperature, surface gravity,
microturbulence velocity and metallicity are consistent (within the uncertainty limits) with the
mean values derived by other authors.

\begin{figure}
\centering
\includegraphics[width=7.5cm,height=7.5cm]{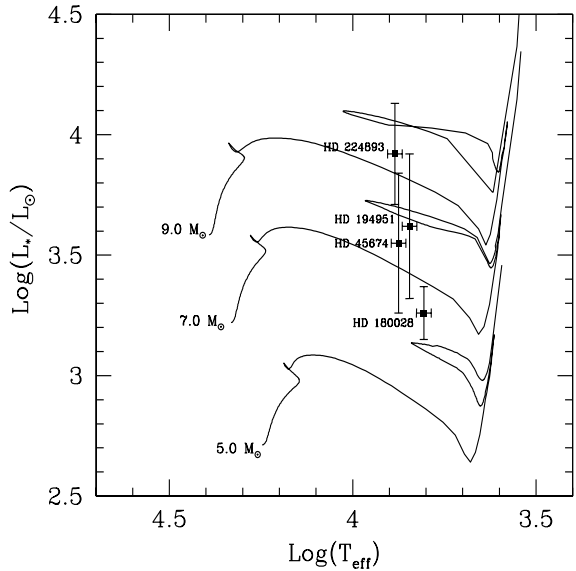}
\caption{Locations of HD\,45675, HD\,180028, HD\,194951 and HD\,224893 on the HR-diagram.
The theoretical evolutionary tracks without rotation with masses of 5, 7 and 9\,M$_{\odot}$ are taken
from Ekstr\"om et al.\@ (2012) at Z\,=\,0.020.}
\label{fig:figure4}
\end{figure}

\subsection{Masses}
\label{sec:mass}

We also estimated the masses of our sample stars. These can be obtained from their
position on the HR-diagram using theoretical evolutionary tracks. We have employed
the theoretical evolutionary tracks without rotation with masses of 5, 7 and 9\,M$_{\odot}$
which are taken from Ekstr\"om et al.\@ (2012) at solar metallicity (Z\,=\,0.020).
Luminosities for HD\,45674, HD\,194951 and HD\,224893 were calculated using
M$_\mathrm{V}$\,=\, V$_\mathrm{0}$ + 5 - log($D$), where V$_\mathrm{0}$ is the intrinsic 
colour derived from
expresion V$_\mathrm{0}$\,=\, $V$ - 3.1\,$E(B-V)$ and $D$ is the distance estimated from parallaxes
(van Leeuwen 2007).

For HD\,180028, however, the absolute magnitude have been derived from a mean value
(M$_\mathrm{V}$\,=\,$-$3.38$\pm$0.16 mag) estimated by Kovtyukh
et al. (2012). The bolometric corrections are taken from Masana et al.\@ (2006) and we adopt
a solar bolometric magnitude, M$_\mathrm{Bol}$\,=\,4.73$\pm$0.01 (Gray 2005).

The colour excess for HD\,180028, HD\,194951 and HD\,224893 was obtained using $E(b-y)$
derived from Str\"omgren photometry by Gray et al.\@ (2001b). We obtain $E(B-V)$ from the
relation $E(b-y)$\,=\,0.78~$E(B-V)$ (Fernie 1987). For HD\,45674 the colour excess is estimated
from Schlafly's map (Schlafly \& Finkbeiner 2011) and where the dust in the Galactic disc is
modelled assuming a thin exponential disc with a scale-height of 125\@ pc. A correction
to colour excess $(B-V)$ have been
done for the last assumption. Their values are present at the eleventh column
in Table~\ref{tab:table1}.

Figure~\ref{fig:figure4} shows the position of our sample on the HR-diagram.
The uncertainty in the luminosity is based in the uncertainties of the parallaxes,
visual magnitudes, extinctions and bolometric corrections. Individually, these uncertainties
varies like 0.29 (HD\,45674), 0.11 (HD\,180028), 0.30 (HD\,194951) and 0.21 (HD\,224893)
respectively. According to this position we could to point out that HD\,180028 has
a mass that vary from 5 to 7\,M$_{\odot}$, HD\,45674 and HD\,194951 between
7-7.5\,M$_{\odot}$ and HD\,224893 of $\sim$9\,M$_{\odot}$.

\subsection{Evolutionary status}
\label{sec:evolutionary}

CNO abundances are key to deduce the evolutionary status since their composition
reflects the mixing process inside of the stars. These are found summarized in
Table~\ref{tab:table4}. With respect to the light elements CNO, we can observe that C
abundance for HD\,45674, HD\,180028, HD\,194951 and HD\,224893 shows deficiency
under non-LTE correction, i.e., these range from $-$0.04 to $-$0.69 dex.

For the total sample we were able
to identify only one nitrogen line ($\lambda$4145.4\,\AA\@) for HD\,45674 and HD\,224893.
Their values under NLTE corrections are [N/Fe]\,=\,$+$0.36$\pm$0.11 dex
and [N/Fe]\,=\,$+$0.99$\pm$0.14 dex respectively. Both objects show signs of CN processed
material that have
been released to the surface. Being unable to estimate the abundance of nitrogen in HD\,180028,
we can not argue about the eficiency of the CN-cycle on its surface.
On the contrary, the N abundance in HD\,194951 has been previously studied by
Lyubimkov et al.\@ (2011). In fact, Lyubimkov et al.\@ (2011) found that the nitrogen is
enriched, i.e., [N/Fe]\,=\,$+$0.73$\pm$0.18 dex.

HD\,45674, HD\,194951 and HD\,224893 show an enhancement of the N abundance.
These objects have shown a systematic deficience in the C abundance and a systematic
increase in the N abundance indicating the presence of CN-cycle material in stellar surfaces.
This enhancement of nitrogen have been also
reported previously by Venn (1995b), Venn \& Przybilla (2003), Luck \& Lambert (1985),
Luck \& Wepfer (1995) and Smiljanic et al.\@ (2006) in supergiant and bright giant stars.

Three process have been suggested as a result of the surface nitrogen enhancement
for massive stars. Consequently, each of case leads to the presence of CN-cycle material
in stellar surfaces. In single stars, these process can be due to the severe mass loss,
the rotational induced mixing during the MS phase and the dredge-up associated with the
deep convective envelope. According to Lyubimkov et al.\@ (2011) the mass loss is one of
the process to be unimportant for B-type stars (progenitors of A--F supergiants) within
of mass range between 4--15\,M$_{\odot}$. By contrast, the other two process might be found
coupled, e.g., it is an observational fact that the mixing process during the 1DUP event might
differ between rotating and non-rotating stars. In binary stars, on the contrary, this
enhancement is subject to mass transfer of its companion evolved.

In order to verify whether HD\,45674, HD\,194951 and HD\,224893
have already passed through the 1DUP event
is necessary to know the [N/C] ratio. The post-1DUP prediction done by Schaller et al.\@ (1992) 
is approximately [N/C]\,=\,$+$0.60 dex for stars between 2 and 15\,M$_{\odot}$.
On the other hand, Meynet \& Maeder (2000)
calculate evolutionary rotating (with a initial velocity of 300\,km\,s$^{-1}$) and non-rotating
models for masses between
9-120\,M$_{\odot}$. These authors predict [N/C]\,=\,$+$0.72 dex without rotation and
[N/C]\,=\,$+$1.15 dex with rotation after the 1DUP and during the blue loop phase for stars
with 9\,M$_{\odot}$. For more details see Figure 14\@ in Smiljanic et al.\@ (2006).

\begin{figure}
\centering
\includegraphics[width=7.5cm,height=7.5cm]{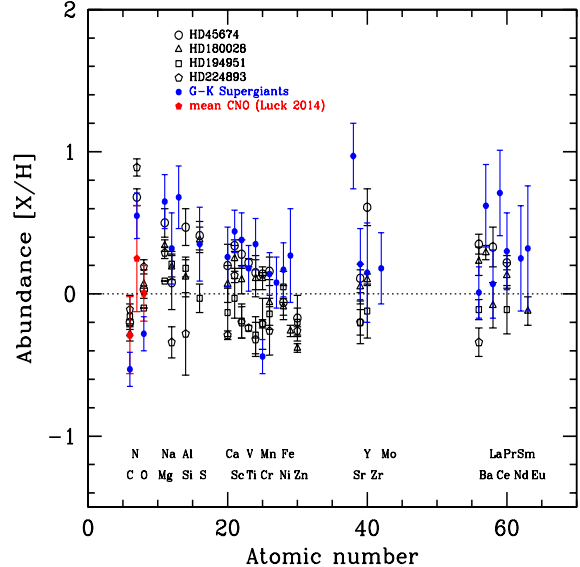}
\caption{Comparison of our abundances [X/H] with the mean abundances derived for a sample
of G--K young supergiants taken from Luck (1977; 1978) and mean CNO abundances by Luck (2014).
Symbols and colour represent data from different sources.}
\label{fig:figure5}
\end{figure}

For HD\,194951, the [N/C] ratio could be obtained from N abundance derived by
Lyubimkov et al.\@ (2011) and C abundance estimated in this work.
Here, we must emphasize that our model differs 240\,K in T$_\mathrm{eff}$ and 0.12\@ in log\,$g$ from
the model used by Lyubimkov et al.\@ (2011). From Table~\ref{tab:table3} we can observe
that there are small differences of $-$0.02 (in T$_\mathrm{eff}$) and $-$0.06 (in log\,$g$)
respectively. These small differences does not affect
the N abundance derived by Lyubimkov et al.\@ (2011) due to a change of model.
Hence, it is possible to use the nitrogen abundance derived by Lyubimkov et al.\@ (2011)
with our C abundance. This value is [N/C]\,=\,$+$0.81$\pm$0.17 dex.
For HD\,45674 and HD\,224893 we have determined a [N/C] ratio of $+$0.88$\pm$0.09 dex and
$+$1.00$\pm$0.13 dex respectively. We note also that these values indicates that both stars show signs
of an efficient mixing process. The mean [N/C] ratio for the sample stars is
[N/C]\,=\,$+$0.90$\pm$0.13 dex.

In short, our results for single and mean [N/C] ratio are larger than
the value obtained by Schaller et al. (1992) by a factor between 0.3-0.4 dex
and are slightly above of the non-rotating model predicted by Meynet \& Maeder (2000), i.e.,
between $+$0.72 dex (non-rotating model) and $+$1.15 dex (rotating model).
Our mean value $+$0.90$\pm$0.10 dex would be in agreement with the results for
non-rotanting stars with masses of 9\,M$_{\odot}$.
Therefore, we may conclude that within of the range of masses that cover our objects
(5--9\,M$_{\odot}$) and their [N/C] ratios, HD\,45674, HD\,194951 and HD\,224893 have
already experienced the 1DUP event and are in post-1DUP phase.
Evolutionarily, the post-1DUP objects can be located only inside
the red-blue loop area. Claret (2004) suggests that extensive red-blue loops
occur for stars with masses from 6 to 13\,M$_{\odot}$, although this extension is still
a matter of controversy (see \S6 and Figure 12 in Lyubimkov et al.\@ 2011).

On the other hand, if we consider the effective temperature and luminosities to be right,
the position of HD\,45674, HD\,194951 and HD\,224893 appears to be located within
the extended red-blue region with $M$\,=\, 7--9\,M$_{\odot}$ on the HR-diagram. However,
the uncertainties causing an appreciable change in the luminosity so that its evolutionary
status is unreliable. With regard to HD~180028 if we consider it to have a mass of
5\,M$_{\odot}$ its probable evolutionary status would be located in the red-blue loop region
(post-1DUP phase), on the
contrary, if we consider it with a mass of 7\,M$_{\odot}$ it appears to be located on the Hertzsprung
gap region and moving for the first time towards the red giant/supergiant phase (post-MS phase).

Under non-LTE correction HD\,45674, HD\,180028, HD\,194951 and HD\,224893
shows values in O abundance of $-$0.36, $+$0.02, $-$0.10 and $+$0.53
respectively. HD\,224893, by contrast, has a moderate enhancement ([O/Fe]\,=\,$+$0.45) probably
due to the CNO bi-cycle.
In general, our oxygen values show the observed tendency in objects of the disc studied by
Bensby et al.\@ (2014); da Silva et al.\@ (2015).

\subsection{Sodium element}
\label{sec:na}

In our sample we found a moderate enrichment of Na abundances, i.e., [Na/Fe] of
$+$0.34, $+$0.43, $+$0.24 and $+$0.55 derived for HD\,45675, HD\,180028,
HD\,194951 and HD\,224893 respectively.
It is believed that the sodium enrichment is related to the
first dredge-up event (Denissenkov \& Denissenkova 1990)
although this assumption is questioned. The predictions made by models (see Fig. 8\@ in  Karakas
\& Lattanzio 2014) indicate that the abundance of sodium does not show a significant
enrichment in low-mass stars ($M$\,$\le$\,2\,M$_{\odot}$) from the first dregde-up or by extra-mixing
process to solar metallicity (Charbonnel \& Lagarde 2010). Even for $M$\,$>$\,2\,M$_{\odot}$
this enrichment does not exceed 0.3 dex.

A non-LTE correction about of $-$0.10 dex to Na was taken from Lind et al.\@ (2011) and
involves values of [Na/Fe] of $+$0.16, $+$0.22, $+$0.24 and $+$0.37 dex  respectively.
We noted that our observational results (within their uncertainties) are in agreement with
[Na/Fe] predicted by El Eid \& Champagne (1995) for intermediate mass stars
($\sim$0.2-0.3 dex).

\subsection{Heavier elements}
\label{sec:heaver}

The tendency of alpha-elements for our objects is similar to that observed
in the thin disk population, i.e., [$\alpha$/Fe] range from $-$0.02 to 0.18 dex.
The sulfur abundance was estimated synthesizing the spectral region at (6743--6757\,\AA\@).
[S/Fe] of $+$0.25 for HD\,45675 and $+$0.12 for HD\,194951
shows a similar tendency observed to $\alpha$-elements of the disc to
metallicity near to solar value (Caffau et al.\@ 2005). By contrast, S abundance for HD\,180028
is overabundant, i.e., [S/Fe]\,=\,$+$0.44 dex. Iron-peaks and neutron-capture elements in all
sample stars show [X/Fe] abundances with tendency towards to solar value. HD\,45674 and
HD\,180028 shows a modest enrichment of \ion{Zr}{ii} and \ion{La}{ii} of $+$0.45 dex and $+$0.32 dex
respectively.

\subsection{Comparison with other supergiants}
\label{sec:comparison}

Our [X/H] abundances can be compared with the mean abundances derived from a
sample of G--K young supergiant stars studied by Luck (1977; 1978) and mean CNO abundances by
Luck (2014).
Figure~\ref{fig:figure5}
shows the [X/H] abundances versus atomic number. The abundances in the sample stars of the
present paper are shown with different symbols together with their uncertainties, i.e., HD\,45674
(open circle), HD\,180028 (open triangle), HD\,194951 (open square) and HD\,224893 (open pentagon),
the cool supergiants are represented with a blue filled circle and CNO abundances with red
filled pentagon respectively.
We can see, in general, that abundances of our sample stars (e.g., CNO, $\alpha$, Fe-peak and
neutron capture
elements) follow the trends expected from Galactic chemical evolution of the Population I.

\section{Conclusions}
\label{sec:concluss}

From this study we performed a detailed analysis of the photospheric abundances for
a sample of four A--F type supergiant stars of intermediate mass ($\sim$5--9\,M$_{\odot}$) using
high-resolution spectra. We have determined
atmospheric parameters, masses and abundances using spectral synthesis and equivalent widths.
Our three stars (HD\,45674, HD\,194951 and HD\,224893) for which we were able to determine both carbon and
nitrogen show signs of internal mixing. A mean [N/C] ratio of $+$0.90$\pm$0.13 dex is found and
this value is in agreement with the preditions of non-rotating models by Meynet \& Maeder
(2000), which predict [N/C]\,=$+$0.72 dex.
A surface nitrogen enhancement was observed in these stars and it only could comes from as a
result of deep mixing during the 1DUP.
The sample of three stars shows very little variability in radial velocities by discarding binarity
in them.

According to abundances analysis we can conclude that HD\,45674, HD\,194951 and HD\,224893
show typical values of abundances for supergiants of the thin disc population and where
evolutionarily the post-first dregde-up (post-1DUP) phase was reached. These objects are located
in the red-blue loop region. HD\,180028, on the contrary, also show typical abundances of
the population I but its evolutionary status could not be satisfactorily defined.

\section*{Acknowledgments}

This work has made extensive use of ELODIE, SIMBAD and ADS-NASA database to which we are
thankful. The authors thanks an anonymous referee for fruitful comments and suggestions to
improve this work.

\end{document}